%% file: main.arXiv.tex
\documentclass[reprint, amsmath, amssymb, superscriptaddress,showkeys,aps,pra]{revtex4-2}

\usepackage{imports}
\usepackage{natbib}
\begin{document}

\title{Quantum ground-state cooling of two librational modes of a nanorotor}
\input{authors}

\begin{abstract}
Controlling the motion of nanoscale objects at the quantum limit promises new tests of quantum mechanics and advanced sensors. Rotational motion is of particular interest, as it follows nonlinear dynamics in a compact, closed configuration space, giving rise to a plethora of phenomena and applications beyond the possibilities of free or trapped linear motion.
A prerequisite for such experiments is the capability to trap nanorotors and initialize them in a quantum ground state of libration.
Here, we demonstrate the reliable, repetitive laser-induced loading of silica nano\-dimers and trimers into an optical tweezer.
Coherent scattering in a high-finesse cavity allows us to cool two different librational modes to the quantum ground state with occupation numbers as low as $n_{\beta}=0.54\pm0.32$ and $n_{\alpha}=0.21\pm0.03$. 
By simultaneously cooling both degrees of freedom ($n_\beta=0.73\pm0.22$, $n_\alpha=1.02\pm0.08$) we align nanorotors to a space-fixed axis with precision better than \qty{20}{\micro rad}, close to the zero-point amplitude of librations.
\end{abstract}


\maketitle

While the harmonic oscillator is one of the first examples of introductory quantum physics \cite{Sakurai_Napolitano_2020} and the central paradigm of quantum nanomechanics \cite{delic_cooling_2020,tebbenjohanns_quantum_2021,magrini_real-time_2021,ranfagni_two-dimensional_2022,deplano2025high,piotrowski_simultaneous_2023,wang2024ground,whittle2021approaching,teufel2011sideband}, quantum rotors are only starting to receive the attention they deserve~\cite{hoang_torsional_2016,Stickler2018,Rashid2018,koch_quantum_2019}. Still in the classical domain, nanoscale rotors have recently been pushed to new experimental limits: Dielectric nanorotors were driven to rotation rates of several GHz~\cite{Ahn2018,reimann_ghz_2018, Yu2021, Ahn2020,jin_6_2021}, exploring material stress limits and achieving high torque sensitivity. Such experiments pave the way for future tests of vacuum friction and nanoscale magnetism or the search for non-Newtonian forces near surfaces~\cite{Schuck2018,ju_near-field_2023}. A nanorod was also shown to serve as a nanomechanical hand of a clock and a pressure sensor with micrometer spatial resolution~\cite{kuhn_optically_2017}.

In the quantum regime, rotational dynamics are intriguing because they are non-linear and periodic in a compact phase space~\cite{goldstein2002classical}, which gives rise to new phenomena, such as persistent quantum tennis racket flips~\cite{ma_quantum_2020}. At the same time, the strong coupling between spin and mechanical angular momentum through the Einstein-de Haas effect offers many opportunities for creating and observing mechanical quantum states \cite{yin2013large,hoang2016electron,rusconi_quantum_2017,Delord2018,delord_spin-cooling_2020,arita2023cooling,Jin2024}.
Prospects for interference experiments with dielectric nanorotors are particularly interesting: In contrast to linear matter-wave interferometry, rotational wave functions can divide up and recombine naturally,  without the need for any beam splitter or mirror and even for a particle pinned in real space~\cite{stickler_quantum_2021}. This opens new windows for the preparation of high-mass rotational Schrödinger cat states~\cite{stickler_probing_2018}, which would also be sensitive to models of wave function collapse~\cite{Schrinski2017} or even dark matter~\cite{Riedel2017,carney_mechanical_2021,Moore_2021}.
 
First experimental steps into rotational quantum physics were already taken with small molecules~\cite{Bretislav1995,seideman_revival_1999,poulsen_nonadiabatic_2004}. The main challenge to observing quantum effects with larger objects in the MDa--GDa mass regime lies in being able to trap, cool, and initialize nanorotors in a state that is aligned at the quantum level. The rotor alignment occurs in an external dipole potential between a laser light field and the particle's anisotropic optical polarizability. The nanoparticle rotations then resemble a pendulum that swings harmonically in two dimensions at fixed eigenfrequencies and with limited angular amplitudes. However, in contrast to a classical pendulum, which can take on a continuum of well-defined angles and angular momenta, the quantum rotor is characterized by discrete energy eigenvalues in a harmonic angular potential. It has been proposed that these quantum librations can be cooled to prepare the principal rotor axis with quantum-limited accuracy in its orientation using cavity cooling~\cite{stickler_rotranslational_2016,schafer_cooling_2021,rudolph_theory_2021}. So far, this method has been used to achieve temperatures in the millikelvin range for all librational degrees of freedom of a dielectric nanoparticle~\cite{kamba_nanoscale_2023,pontin_simultaneous_2023}, while cooling to a pure quantum ground state was recently realized for a single librational mode~\cite{dania_high-purity_2025}.

Here, we demonstrate cooling of two librational degrees of freedom deep into their quantum ground states and simultaneously cool both modes such that the nanorotor’s orientation is defined at the quantum limit. We achieve this by actively reducing laser phase noise by three orders of magnitude at both mechanical frequencies, and by exploiting two orthogonal, non-degenerate polarization modes of a high-finesse cavity. Using laser-induced loading of the trap in prevacuum, we achieve unrivaled repetition rates enabling us to cool different nanorotors to their librational quantum ground state, from silica dimers over trimers to clusters, in the same setup in a single day.

\section{Experimental setup}

\begin{figure}[htbp]
   \centering
    \includegraphics[width=\linewidth]{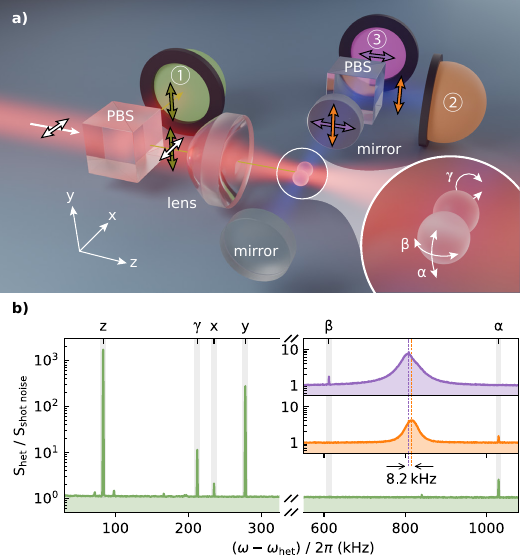}
    \caption{\textbf{Nanorotors trapped in an infrared tweezer.}
        \textbf{a)} A silica nanorotor is trapped in an optical tweezer formed by light propagating along the $z$-axis and polarized along the $x$-axis, and focused with a high-numeric-aperture lens. The $y$-polarized backward-scattered light is collected and monitored on a heterodyne detector \circled{1}. The coherently scattered tweezer light of the nanorotor populates the optical cavity formed by two mirrors. The orthogonally polarized cavity modes are split with a polarizing beam splitter (PBS) and monitored in heterodyne detectors \circled{2} and \circled{3}. Inset: Librational modes $\alpha$, $\beta$, and $\gamma$ in the defined reference frame.
        \textbf{b)} Power spectral densities (PSD) $S_{\text{het}}$ of the heterodyne detection of the backward-scattered light (\circled{1}, green), and cavity modes polarized along the $y$- (\circled{2}, orange) and $z$-axis (\circled{3}, violet) show all degrees of freedom. The spectra are taken at a tweezer–cavity detuning of $\Delta/2\pi \approx \qty{800}{kHz}$, normalized to the shot-noise level $S_{\text{shot noise}}$ and shown here with respect to the heterodyne frequency $\omega_\text{het}$.
        Librations $\alpha$ and $\beta$ are visible in different cavity modes, which are frequency-separated by $\sim\qty{8.2}{kHz}$.
        }
    \label{fig:1-setup_and_two-mode_cooling}
\end{figure}

Our experimental platform is shown in Fig.~\ref{fig:1-setup_and_two-mode_cooling}a. The nanorotors are assembled from two or more silica spheres with a nominal diameter of $d=\qty{119}{nm}$ for the cooling of a single (1D) and \qty{156}{nm} for the cooling of two librational modes (2D). Single spheres, dumbbells, trimers, or clusters are launched at low pressure ($\sim6\,\text{mbar}$) using laser-induced desorption (LID) and trapped in the optical tweezer (wavelength of $\lambda=1550$ nm) propagating along the $z$-axis and linearly polarized along the $x$-axis.  

Although dumbbells and linear trimers are rotationally symmetric about their figure axis, we describe each nanorotor as an asymmetric rigid body with distinct moments of inertia. This accounts for imperfections in individual spheres~\cite{rudolph_theory_2021}, for birefringence~\cite{arita_rotational_2016}, and for limitations of the Rayleigh-Gans approximation~\cite{stickler_rotranslational_2016}.
The orientation of the particle in the space-fixed lab frame $({\bf e}_x,{\bf e}_y, {\bf e}_z)$ is specified by the three Euler angles $(\alpha,\beta,\gamma)$ in the $z$-$y'$-$z''$ convention, which are shown in the inset of Fig.~\ref{fig:1-setup_and_two-mode_cooling}a. The nanorotor aligns its figure axis along the linear polarization of the tweezer, which strongly traps the libration along $\alpha$ and $\beta$ with frequencies in the range of \qtyrange{0.5}{1.3}{MHz} (Fig.~\ref{fig:1-setup_and_two-mode_cooling}b).

The nanorotor is trapped in the center of an optical cavity oriented along the $x$-axis, with a finesse of $\mathcal{F}\approx300\,000$, an energy decay rate of $\kappa/2\pi=\qty{32.4}{kHz}$, and a detuning $\Delta =\omega_\mathrm{c}-\omega_\mathrm{l}$ with respect to the tweezer frequency $\omega_\mathrm{l}=2\pi c/\lambda$ (Fig.~\ref{fig:1-setup_and_two-mode_cooling}a). The cavity is slightly birefringent and its eigenmodes are polarized along the $y$- and $z$-axis with a frequency splitting of about $8.2\,\text{kHz}$. The nanorotor scatters the tweezer light into an initially empty cavity mode, thereby coupling its motion to the cavity field. 
The coupling is described by the interaction potential
\begin{equation}\label{eq:quantinteraction_hamiltonian}
    \frac{U_{\rm int}}{\hbar} =  (g_\alpha a_y + g_\alpha^* a_y^\dagger) (b_\alpha +b_\alpha^\dagger) + (g_\beta a_z + g_\beta^* a_z^\dagger) (b_\beta + b_\beta^\dagger),
\end{equation}
where $b^{(\dagger)}_{\mu}$ are the librational ladder operators and $g_{\mu}$ the respective coupling rates with $\mu\in\{\alpha,\beta\}$ (see Suppl. Inf.) and $a^{(\dagger)}_{\nu}$ are the ladder operators associated with the cavity modes with $\nu\in\{y,z\}$.
Remarkably, the two librations couple selectively to orthogonal cavity polarizations: the $\alpha$-libration to the y-polarized mode $a_y$, and the $\beta$-libration to the z-polarized mode $a_z$.
The total Hamiltonian then describes the energy associated with the population of the two cavity modes, the two mechanical modes of frequency $\Omega_\mu$ as well as the interaction between them:
\begin{align}\label{eq:hamiltonian}
    H  = \sum_{\nu =y,z}\hbar\Delta a_\nu^\dagger a_\nu + \sum_{\mu = \alpha,\beta}\hbar\Omega_\mu b_\mu^\dagger b_\mu + U_{\rm int}
\end{align}

To unambiguously identify the two librational modes $\alpha$ and $\beta$, we use the discrimination of the cavity eigenmodes by polarization.
We therefore split the transmitted cavity light into its eigenmodes $a_y$ (detector \circled{2}) and $a_z$ (detector \circled{3}). Each signal is mixed with a local oscillator detuned by $\omega_{\text{het}}$ yielding the heterodyne signal displayed in Fig.~\ref{fig:1-setup_and_two-mode_cooling}b (see Suppl. Inf., Fig.~\ref{fig:ext-1-setup}c).
According to the theoretical coupling, peaks corresponding to $\alpha$ ($\beta$) appear in the orange (violet) power spectral density (PSD) of $a_y$ ($a_z$).
For the SiO$_2$ cluster trapped in Fig.~\ref{fig:1-setup_and_two-mode_cooling}b, we identify peaks with frequencies of $\Omega_\beta=2\pi\times\qty{612}{kHz}$ and $\Omega_\alpha=2\pi\times\qty{1030}{kHz}$.
Since cavity transmission is affected by cavity-enhanced laser phase noise (see Suppl. Inf., Fig.~\ref{fig:ext-2-noise}c), we use the heterodyne detection of the backward-scattered tweezer light (detector \circled{1}) to monitor all degrees of freedom.

When the cavity is blue-detuned relative to the tweezer light ($\Delta>0$), the interaction Hamiltonian \eqref{eq:hamiltonian} predicts cooling via coherent scattering. At resonance, the cavity enhances the spectral mode density.  If its detuning to the tweezer matches the level spacing of the mechanical harmonic oscillator, the probability of anti-Stokes scattering is increased. In this process, the librational oscillator is transferred from quantum number $n$ to the next-lower level $n-1$, while the  photon energy is increased from $\hbar\omega_l$ to $\hbar(\omega_l+\Omega_\mu)$. Repeating this process cools the mechanical motion. In contrast, Stokes scattering raises the oscillator quantum number and reduces the photon energy, thereby heating the nanorotor motion. However, such photons are off-resonant with respect to the cavity and therefore suppressed. On average, this imbalance transfers mechanical energy into photon energy, which exits the cavity~\cite{hechenblaikner_cooling_1998,vuletic_three-dimensional_2001,delic_cooling_2020}.

The minimum phonon occupation $n$ is reached when the cooling rate balances the heating rate $\Gamma$.
The net cooling rate is determined by the difference between the anti-Stokes $A^-$  and Stokes $A^+$ scattering rates, while additional heating comes from collisions with background gas and photon recoil ~\cite{jain_direct_2016}. The equilibrium phonon occupation of the librational modes is therefore given by:
\begin{align}
    n_\mu=\frac{\Gamma_\mu+A_\mu^+}{A_\mu^--A_\mu^+} +n_\phi(\Omega_\mu)
    \label{eq:final_occupation}
\end{align}
The phase noise contribution $ n_\phi(\Omega_\mu) \approx S_\phi(\Omega_\mu) n_{\text{cav}}/\kappa$ is determined by laser phase noise $S_\phi(\Omega_\mu)$ and the intracavity photon number $n_{\text{cav}}$~\cite{rabl_phase-noise_2009,Meyer2019}. For our laser, this would prevent us from reaching ground-state cooling. To mitigate it, we measure the phase noise using an unbalanced Mach-Zehnder interferometer (Fig.~\ref{fig:ext-1-setup}b). We reduce its level at the eigenfrequencies of both the $\alpha$- and the $\beta$-librations using active feedback via an electro-optical modulator (see Suppl. Inf., Fig.~\ref{fig:ext-2-noise}b--c) \cite{parniak_high-frequency_2021,dania_high-purity_2025}. The role of gas collisions is reduced by pumping the chamber to about \qtyrange{2.5e-8}{3.5e-8}{mbar}.

\section{One-dimensional ground-state cooling}

\begin{figure}[htbp]
   \centering
    \includegraphics[width=1\linewidth]{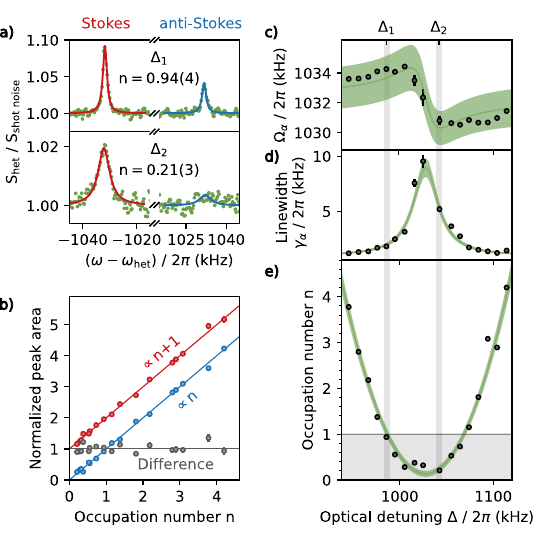}
    \caption{\textbf{1D ground-state cooling.} \textbf{a)} PSD of the Stokes (red) and anti-Stokes scattering (blue) for $\Delta_1/2\pi=\qty{986}{kHz}$ and $\Delta_2/2\pi=\qty{1042}{kHz}$, overlaid for comparison. The imbalance of the peak heights is used to extract the occupation $n$. \textbf{b)} The Stokes (red) and anti-Stokes scattering rates (blue) for different occupations $n$ show their proportionality to $n+1$ and $n$. \textbf{c)-e)} The fitted mechanical frequency, linewidth, and the obtained occupation (black points) are shown as a function of the tweezer-cavity detuning $\Delta$. The theoretical fits to these dependencies account for errors of linewidth, coupling, offset-frequency, heating rate and phase noise occupation, which are shown as green regions. The individual errors of most data points are smaller than the marker size.
    }
\label{fig:2}
\end{figure}

We record the cooling behavior specifically of the $\alpha$-libration as a function of the detuning $\Delta$ around $\Omega_\alpha$.  
In Fig.~\ref{fig:2}a, we plot the power spectral densities (PSD) of this motion at positive and negative frequencies to perform sideband thermometry with the Stokes (red) and anti-Stokes (blue) sidebands. We fit the peaks with a Lorentzian profile to extract their frequencies, linewidths, and background noise levels. 

The Stokes and anti-Stokes peak areas $A_{\mathrm{S}}\propto A^+$ and $A_{\mathrm{aS}}\propto A^-$ scale with the occupation number $n$ as $A_{\mathrm{S}} = C~ (n+1)$ and $A_{\mathrm{aS}} = C~n$, with a proportionality factor $C$. The occupation number $n$ can then be extracted from the sideband amplitudes (see Suppl. Inf.).

Already at a cavity detuning of $\Delta/2\pi=986\,\text{kHz}$ the imbalance between the red and blue sideband is 
significant, signaling a phonon occupation of $n=0.94\pm0.04$. Shifting the detuning to $\Delta_2/2\pi=1042\,\text{kHz}$ leads to an occupation of the $\alpha$-libration as low as $n=0.21\pm0.03$, populating the harmonic oscillator quantum ground state with a high probability of $83\pm2\,\%$. 

\begin{figure*}[htbp]
    \centering
    \includegraphics[width=1\textwidth]{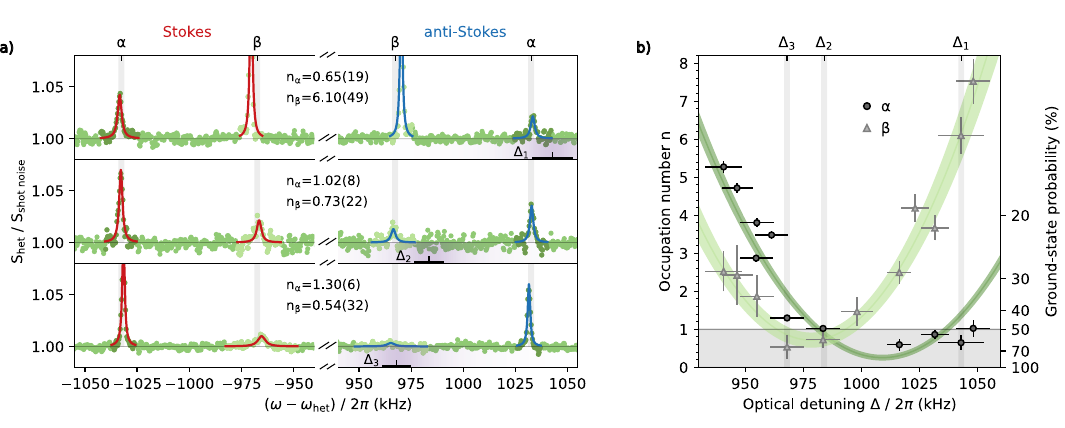}   
    \caption{\textbf{Ground-state cooling of the $\alpha$- and $\beta$-libration.}  \textbf{a)} PSDs of the detector signal \circled{1} (green points) show cooling to the ground state of $\alpha$ (top) and $\beta$ (bottom), and close to the simultaneous ground state of $\alpha$ and $\beta$ (middle), for three different cavity detunings $(\Delta_1,\Delta_2,\Delta_3)=2\pi\times (1043,984,968)$\,kHz. The cavity transfer function is indicated by the purple gradient at the bottom of each plot, with the center frequency and standard deviation shown as a line marker. We extract the mode occupations from the imbalance between the fitted Stokes (red) and anti-Stokes scattering (blue). \textbf{b)} Extracted phonon occupations of the $\beta$- (gray triangles) and $\alpha$-librations (black circles) as a function of the tweezer-cavity detuning $\Delta$. Theoretical predictions for occupations of the $\alpha$- and $\beta$-modes are shown as light and dark green regions, respectively.
    }
    \label{fig:3}
\end{figure*}
 
In Fig.~\ref{fig:2}b, we show the normalized peak areas as a function of the occupation number. As expected for a quantum harmonic oscillator,  the probability for Stokes and anti-Stokes scattering scale with the occupation like $n+1$ and $n$ respectively. The good stability and repeatability of the experiment are seen when we plot the difference (gray) of the normalized side bands for different cavity detunings and hence occupation numbers. It remains consistently close to $1$.

Scanning the cavity detuning around the mechanical resonance changes the optomechanical coupling and therefore also both the eigenfrequency (Fig.~\ref{fig:2}c) and the linewidth (Fig.~\ref{fig:2}d) of the librational motion. Forces due to the intracavity light field act like a frequency-tunable optical spring and damping. Although the frequency change is only on the level of a few per mille, we can clearly resolve it. At the same time, the detuning strongly affects the cooling and final occupation number as plotted in Fig.~\ref{fig:2}e. Its lowest value is the one shown in Fig.~\ref{fig:2}a/Fig.~\ref{fig:2}c, deep in the librational quantum ground state.

Fitting the curves of Figs.~\ref{fig:2}c--d allows us to extract the optomechanical coupling $g_\alpha$ (see Suppl. Inf.). Since the geometry of the nanorotor determines the coupling strength, we can calculate the particle-specific moment of inertia about the $z$-axis as $I_b=3.3\pm0.4\times 10^{-32}\,\text{kg\,m}^2$. In the coldest state, we determine the standard deviation of the rotational mode as $\sigma_\alpha=\qty{17.4\pm0.9}{\micro rad}$, corresponding to a temperature as low as $T_\alpha=\qty{28\pm2}{\micro K}$.

\section{Ground-state cooling of two librational modes}

To tightly align the particle with the polarization axis, we extend coherent scattering cooling now to both modes, $\alpha$ and $\beta$. 

A dumbbell formed from two silica spheres with $d=\qty{156}{nm}$ is trapped and oscillates at $\Omega_\beta/2\pi=978\,\text{kHz}$ and $\Omega_\alpha/2\pi=1035\,\text{kHz}$. As they differ only by about twice the cavity decay rate $\kappa$, they can be cooled simultaneously if the tweezer-cavity detuning is properly chosen and the phase noise suppression is activated at both frequencies (Fig.~\ref{fig:ext-2-noise}c).

\begin{figure}[htbp]
    \centering
        \includegraphics[width=1\linewidth]{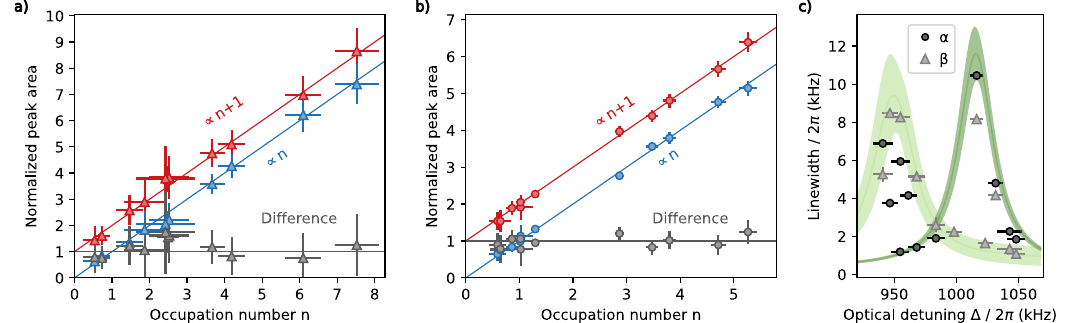}
    
    \caption{{\bf Evaluation of phonon occupation and optomechanical coupling strength for 2D cooling.}
    \textbf{a)--b)} The areas of the Stokes (red markers and linear fits) and anti-Stokes peaks (blue markers and linear fits) are normalized to their respective differences, for both (a) $\beta$ and (b) $\alpha$. Based on the peak areas, we determine the occupation number $n$ (see Suppl. Inf.). The error bars are statistical $1\,\sigma$ values, derived from the measurement of the peak area.
    \textbf{c)} Optical damping due to the cavity coupling: Sweeping the detuning $\Delta$ between the cavity and the tweezer frequency allows to scan across the mechanical resonance of both librational modes. We show the fitted linewidth (green) for the $\beta$-mode (gray triangles) and the $\alpha$-mode (black circles). The green shaded areas account for fitting errors. Due to experimental instabilities of the trapping frequency, some linewidth measurements show excessive values, which were excluded from the fit.
    }
    \label{fig:ext-6-2D}
\end{figure}

We measure the occupation of both librational modes as a function of the detuning $\Delta$ via sideband thermometry (Fig.~\ref{fig:ext-6-2D}a--b). Because the librational modes $\alpha$ and $\beta$ couple to orthogonal cavity modes, we can treat their dynamics separately. By setting the detuning close to the librational frequencies $\Omega_\alpha$ or $\Omega_\beta$, we can cool the $\alpha$- or $\beta$-motion individually into their quantum ground states, $n_\alpha=0.65\pm0.19$ and $n_\beta= 0.54\pm0.32$, as shown in Fig.~\ref{fig:3}a. 
At $\Delta/2\pi=984\,\text{kHz}$, we achieve the lowest combined phonon number  with $n_\alpha=1.02\pm0.08$ and $n_\beta=0.73\pm0.22$. Again, we fit the occupations as a function of detuning (Fig.~\ref{fig:3}b). 
Evaluation of the optomechanical coupling from the oscillator linewidth (see Suppl. Inf., Fig.~\ref{fig:ext-6-2D}c) reveals that the nanorotor is aligned along the $x$-axis with a quantum uncertainty of $\sigma_\alpha=\qty{18\pm1}{\micro rad}$ and $\sigma_\beta=\qty{17\pm3}{\micro rad}$, corresponding to effective librational temperatures of $T_\alpha=\qty{73\pm4}{\micro K}$ and $T_\beta=\qty{57\pm28}{\micro K}$, respectively.
Such cooling at the quantum limit is an important prerequisite for future experiments on rotational interference and quantum sensing~\cite{stickler_quantum_2021}. The aligned state corresponds to a coherent superposition of angular momentum states with mean $j\simeq \sqrt{k_BTI}/\hbar\sim 6\times 10^4$.
If we were to release the rotor non-adiabatically from its orientational ground state, it would evolve into a superposition of rotational quantum states with classically mutually exclusive angular momenta~\cite{Stickler2018}.
This is expected to lead to rotational dispersion and quantum revivals due to the constructive interference of the rotational wave packets after a time $T_\text{rev}=2\pi I/\hbar$. For the nano-dumbbell in our experiment, the revival time would be $50\,\text{min}$. Therefore, observing revivals at a realistically observable timescale requires smaller particles or a scheme to resolve fractional revivals~\cite{Schrinski2022,Ivanov2004}.

\section{Robust and repeatable trap loading and cooling}
\begin{figure}[ht]
    \centering
    \includegraphics[width=1\linewidth]{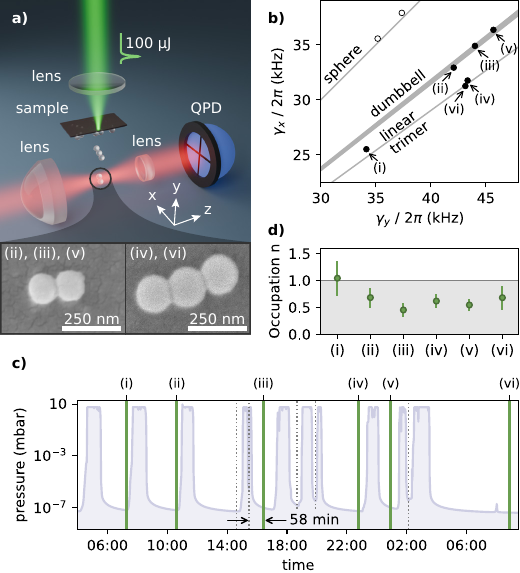}
    \caption{\textbf{Repeatable ground-state cooling.} \textbf{a)} Experimental sketch of loading nanorotors in vacuum with laser-induced desorption (LID). In a scanning electron microscope we observe on a sample prepared with $119\,\text{nm}$-particles besides single spheres and agglomerated clusters also dumbbells and trimers. With a split detection scheme in forward scattering the particle geometry is analyzed by extracting the mechanical damping rates. \textbf{b)} Measured damping rates $\gamma_x$ and $\gamma_y$ for a cluster (i), three dumbbells (ii), (iii), (v), and two linear trimers (iv), (vi), as well as the expected values. \textbf{c)} Pressure trace during loading and ground-state cooling (highlighted by green solid lines). Particles which were ejected are highlighted as gray dotted lines.
    \textbf{d}) Measured occupation numbers of the $\alpha$-libration. The gray-shaded region marks ground state population with a probability greater than $50\%$.}
    \label{fig:4}
\end{figure}

Advanced experiments in levitated optomechanics require sources that can load and cool nanoparticles with high repetition rate and reliability but are at the same time able to handle different particle types and geometries. We demonstrate here the repeatable loading and ground-state cooling of half a dozen different nanorotors, formed from silica nanospheres with $d=\SI{119}{nm}$. We limit this study to the ground-state cooling of the $\alpha$-libration to save measurement time. The particles are coated on a glass slide with a \qty{50}{nm}-thick silicon film on top and placed above the cavity, as shown in Fig.~\ref{fig:4}a. A green laser pulse with an energy of \qty{100}{\micro J} and a duration of about \qty{6}{ns} hits the backside of this sample and ejects the particles into dry nitrogen at a base pressure of \qty{6}{mbar} (see Fig.~\ref{fig:ext-5-closeup} for experimental details). This can release single spheres, dumbbells, trimers, or bigger clusters.
The process is similar to laser-induced acoustic desorption, LIAD~\cite{asenbaum_cavity_2013, Nikkhou2021, bykov_direct_2019}, but because the absorption layer is thin enough to be fully evaporated, we refer to the method as laser-induced desorption, LID.
 As corroborated by scanning electron microscopy, the particles aggregate already in solution, but there is also an indication of the occasional growth of dimers in the trap by sequential capture of two spheres.  

\begin{figure}[htbp]
    \centering
        \includegraphics[width=1\linewidth]{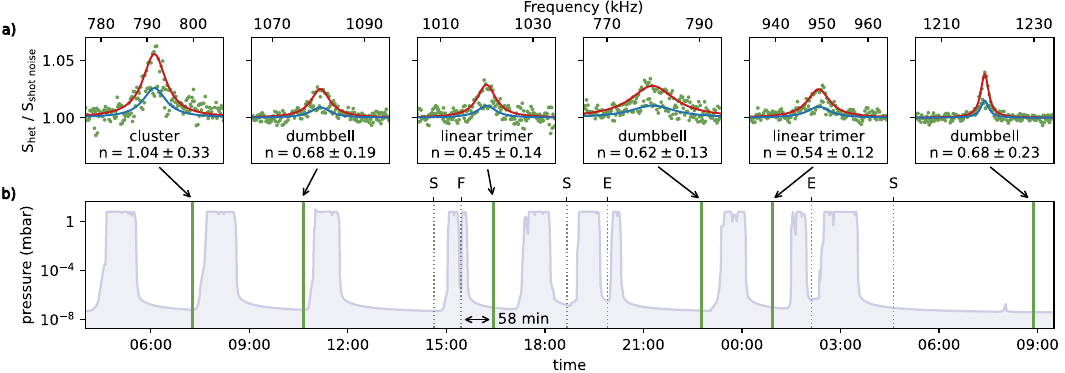}
    \caption{{\bf Ground-state cooling of different nanorotors trapped in the same setup on the same day.}
    {\bf a)} PSDs of the successfully cooled nanorotors, where the red fit corresponds to the Stokes and the blue fit to the anti-Stokes peak. The particles were LID-loaded and characterized at 6\,mbar, evacuated to high vacuum, and cooled by coherent scattering. We demonstrate ground-state cooling of the $\alpha$-motion for three dumbbells and two linear trimers as well as one cluster.
    {\bf b)} The pressure trace during the measurement time acts as a time line, where the successfully cooled particles are marked by vertical green lines. The time stamps of the lab clock show the reliable repetition of the process. The dashed line marked as F indicates that the librational frequency was too low for ground-state cooling. The marker E indicates an operational error. The marker S describes when the $\gamma$-libration heated the other degrees of freedom too much and destabilized the particle orientation. As the last particle demonstrates, ground-state cooling is nevertheless possible after a longer waiting time.
    }
    \label{fig:ext-5-Re}
\end{figure}

To characterize the geometry of the trapped rotor, we track its motion along the $x$- and $y$-axis by monitoring the transmitted tweezer light in a split detection scheme (see Suppl. Inf.). While the oscillator damping in the residual gas is the same along all axes for isotropic nanoparticles, the ratio of the damping rates $\gamma_y/\gamma_x$ amounts to $1.258-1.276$ for dumbbells and to $1.378$ for linear trimers \cite{ahn_optically_2018} (Fig.~\ref{fig:4}b). We have repeated the procedure of trapping, shape assessment, evacuation to high vacuum  and cavity cooling for a series of nanoparticles over a period of $\sim 28$ hours. Half a dozen nanorotors, marked (i)--(vi) in Fig.~\ref{fig:4}b--d, were successfully cooled (near) to their librational quantum ground state. They comprise dumbbells, trimers, and clusters.
To illustrate the scale of such experiments, we plot the trap pressure as a function of time in Fig.~\ref{fig:4}c. Every observation of librational ground-state cooling of a fresh particle is marked by a green line, while gray dashed lines mark events of intentional or accidental particle loss (Fig.~\ref{fig:ext-5-Re}). The fastest cycle from ejection of one particle to ground-state cooling of a new one took $58$ minutes. The final occupation numbers for all six successful events are shown in Fig.~\ref{fig:4}d. 

\section{Conclusion}

Coherent scattering cooling in a high-finesse bimodal cavity enables the simultaneous cooling of two librational modes of a harmonically trapped nanorotor and can define its orientation at the quantum limit. Cooling of a single librational mode is demonstrated  deep in the quantum regime ($n\simeq 0.2$). Laser-induced desorption in prevacuum has opened the path to reliable and repeatable loading and cooling of  several  dumbbells, trimers or clusters within a single day. These advances establish  quantum control over the motional degrees of freedom of nanoparticles with masses around \qtyrange{1}{10}{GDa}, opening new perspectives for real-world quantum force sensing and for tests of fundamental physics. Here we focus on the latter: 

Observing rotational matter-wave interference will require loading, trapping, cooling, and orienting smaller and lighter particles. For a dumbbell made of two $20$\,nm silica spheres, with a total mass of $10$\,MDa, we expect a revival time of about $300$\,ms, corresponding to ca. $40$\,cm of free fall. This would be compatible with a lab-sized free-fall experiment but requires the repeated loading and cooling of particles with very similar properties.
The revival times may also be reduced by considering fractional revivals \cite{Dooley2003,Weber2014}.
Besides the interest in fundamental quantum physics that can test macroscopicity, quantum nanorotors in the range \qtyrange{0.1}{1}{MDa} also have a predicted torque sensitivity of $10^{-30}\,\mathrm{Nm}$~\cite{stickler_probing_2018}, potentially surpassing current sensitivity records by several orders of magnitude. This mass scale is also intriguing because it covers relevant nanobiological materials, from very massive proteins to viruses. The Tobacco Mosaic Virus stands out as a natural nanorotor that can be safely handled and that has very reproducible physical parameters, with \qty{300}{nm} length and \qty{18}{nm} diameter, weighing $40$\,MDa~\cite{Butler1984}. However, such thermolabile bionanomaterials will require indirect cooling methods that leave the particles in the dark~\cite{dago_stabilizing_2024}.


\section*{Acknowledgments}
The authors thank Klaus Hornberger, Iurie Coroli, and Nancy Gupta for useful discussions during the planning and construction of the experiment and Stephan Puchegger for support in SEM imaging.
We thank the nanorotors "Neal Particleton" and "Quinn Tumman", who stayed in the trap for many days, for providing the first impressive data on librational cooling in this experiment and all other nanorotors for their repeated visits.
\section*{Declarations}
\noindent
\textit{Funding:}
The experiment was supported by the US Office of Naval Research award No. N62909-23-1-2029.
ST acknowledges support from the Austrian Academy of Sciences (ÖAW) through an ESQ discovery project.
BAS acknowledges funding by the Carl-Zeiss Foundation through the project QPhoton and by the DFG--510794108. 
UD acknowledges the support of the Austrian Science Fund (FWF) START grant 10.55776/STA175.\\

\noindent
\textit{Author contributions:}
ST conceived the experiment with support from FF, LH, MA, and UD, based on theoretical predictions from HR and BAS. ST, FF, and LH built the experiment. ST and FF acquired and analysed the data. BAS, UD, and MA supervised the project. ST, FF, BAS, UD, and MA wrote the manuscript. All authors were involved in reviewing the draft.
  

\bibliographystyle{naturemag}
\bibliography{compactbib}


\onecolumngrid
\newpage
\section*{Supplementary information}

\subsection{Optical Setup}
\begin{figure}[htbp]
    \centering
    
        \includegraphics[width=146mm]{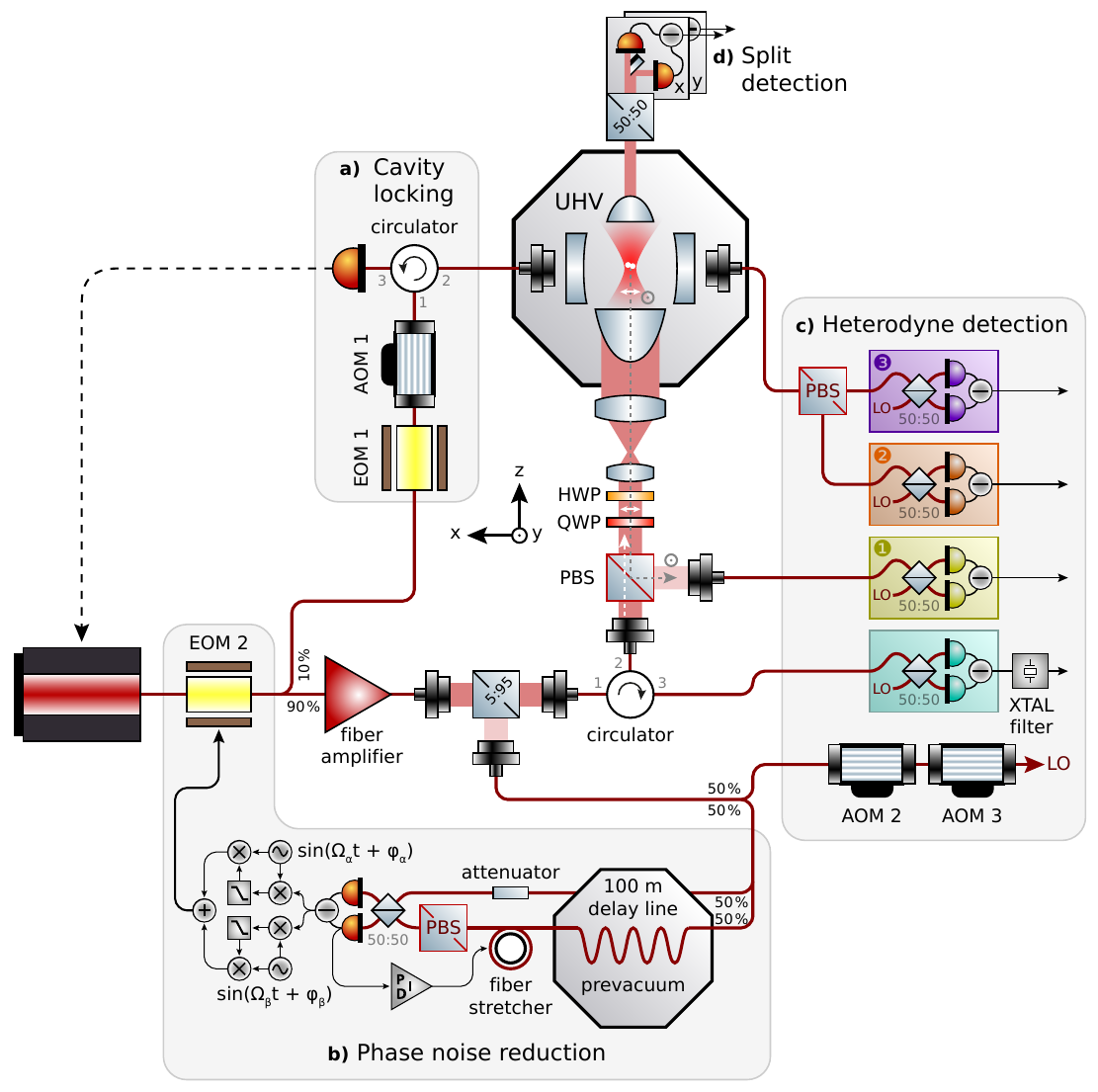}
    
    \caption{{\bf Extended optical setup for trapping, cooling and detection.} The fiber laser beam is modulated by the electro-optical modulator EOM~2 to reduce its phase noise level before it is amplified. The trapping beam is polarization-controlled by a quarter- (QWP) and half-wave plate (HWP), expanded and re-focused into a sub-micron waist at the center of an antinode of the cooling cavity.  {\bf a)} Pound-Drever-Hall locking of the laser to the cooling cavity, shifted by about one free spectral range, i.e. by \qty{9.72}{GHz} (shifted by EOM~1 and acousto-optical modulator AOM~1). The cavity frequency is additionally blue-detuned with respect to the tweezer light by about one mechanical eigenfrequency of the libration that should be cooled, i.e., by \qtyrange{0.5}{1.3}{MHz}.
    {\bf b)} Unbalanced Mach-Zehnder interferometer with \qty{100}{m} fiber loop to measure laser phase noise, in particular sensitive at the librational frequencies around \qty{1}{MHz} off the infrared carrier frequency. A fiber stretcher is used to stabilize the length of both arms. The recombined arms are monitored on a balanced detector, the output of which is used to actively compensate the phase noise at both librational frequencies using EOM~2. 
    {\bf c)} Polarization-sensitive heterodyne detection. Part of the laser light gets up- and down-shifted by AOM~2 and AOM~3 to prepare the local oscillator (LO) for heterodyne detection about \qty{5}{MHz} blue-detuned with respect to the tweezer light.
    Heterodyne detection \circled{1} of the $y$-polarized light in the back-plane is used to monitor all motional degrees of freedom of the nanorotor using the light that is backscattered by the particle, which is mixed with the LO on the photo-detector. An additional heterodyne detector is used to monitor the $x$-polarized light via a fiber circulator. A notch filter based on a crystal oscillator (XTAL) is used to reduce the signal level of Rayleigh scattered light.
    The cavity transmission is used to distinguish both librational modes $\alpha$ and $\beta$ and the cavity mode frequencies on balanced detectors \circled{2} and \circled{3}.
    {\bf d)} The transmitted tweezer light is collimated and halved by a beam splitter. The two halves are fed to two split detection schemes; one sensitive to the motion along the $x$-axis, one along the $y$-axis.
    }
    \label{fig:ext-1-setup}
\end{figure}

The extended optical setup is shown in Fig.~\ref{fig:ext-1-setup}. 
Light emitted by an infrared fiber laser (NKT Photonics Koheras Adjustik E15) passes through the fiber electro-optical modulator EOM~2. 
We split off a small fraction of the light to lock the cavity (Fig.~\ref{fig:ext-1-setup}a). The rest is amplified to a power of $6\,\text{W}$ (NKT Photonics Boostik HP) and then divided into three parts:  one for phase-noise detection, one serving as the local oscillator (LO) in the heterodyne detection (Fig.~\ref{fig:ext-1-setup}c), and up to $3\,\text{W}$ for the optical tweezer.

The tweezer mode is cleaned in a polarization-maintaining fiber, and its polarization is set by wave plates to be linear along the cavity axis. This orientation minimizes Rayleigh scattering into the cavity when the nanorotors are perfectly aligned. The laser light fills the aspherical tweezer lens, which has a 
diameter of $25.4\,\text{mm}$, a numerical aperture of $\text{NA} =0.81$, and an effective focal length of $13.2\,\text{mm}$ (Thorlabs, custom design). For a cluster assembled of $119\,\text{nm}$ nanospheres (Fig.~\ref{fig:1-setup_and_two-mode_cooling}b), we determine a trap power of $P=2.7\,\text{W}$ and the trapping waists of $w_x=\qty{1.17}{\micro m} \mathrm{\,and\,}  w_y=\qty{0.98}{\micro m}$.

We detect the trapped nanoparticle by collecting its back-scattered light, as shown in Fig.~\ref{fig:ext-1-setup}c. Its two polarization components are split by the polarizing beam splitter (PBS) and detected separately. The vertical component provides most information about the particle's rotation, in particular about the rotation around the tweezer propagation axis $z$. This signal is only weakly sensitive to Rayleigh scattering of the aligned rotor and scattering at surfaces along the beam path. Therefore, this component is used to monitor cooling to the librational ground state.
The horizontal contribution is isolated using a fiber circulator, which provides intrinsic alignment of the backscattering signal and is therefore used during trap alignment. In order to reduce the Rayleigh scattering peak, we filter the electrical signal by a crystal oscillator.

\begin{figure}[htbp]
    \centering
    \includegraphics[width=0.75\textwidth]{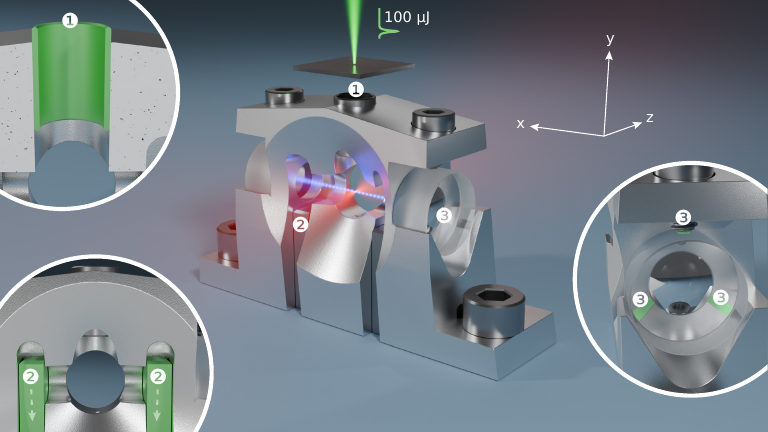}
    \caption{
         {\bf Experimental details.} A pulsed nanosecond laser ($E_\text{pulse}=\qty{100}{\micro J}$) is focused onto a cover slip coated with a \qty{50}{nm} thick layer of silicon and SiO\textsubscript{2} nanoparticles (SP). 
         The particles diffuse in \qty{6}{mbar} of dry nitrogen to the focus of the optical tweezer where they are trapped.
         Over thousands of shots, charge can also accumulate and electrostatically clog the upper hole of the mount, reducing the efficiency of trap loading. A simple remedy to this is a metallic tube \circled{1} in the mount that can be easily exchanged.
         In order to shield the cavity mirrors from contamination they are protected by a retractable shield \circled{2} during the loading process. 
         The cavity mirrors are glued onto the Invar holder. A screw with a soft tip applies an adjustable force on the top of the mirror \circled{3}. Together with contact areas separated by 120\textdegree, a controlled birefringence is induced in the cavity mirror coating, which leads to a cavity mode splitting along the $y$- and $z$-axis.
    }
    \label{fig:ext-5-closeup}
\end{figure}

The trapped nanoparticle is centered at an antinode of the cooling cavity mode. The resonator is formed from mirrors with intensity reflectivity  $R\geq0.99999$ (Five Nine Optics) and radius of curvature $\text{RoC} =5\,\text{cm}$, 
yielding a finesse of $\mathcal{F}=300\,000$ at a free spectral range of $\text{FSR} =9.72\,\text{GHz}$, a linewidth of $\kappa/2\pi=32.4\,\text{kHz}$ and a central waist of $w_\text{cav}=\qty{94}{\micro m}$ (Fig.~\ref{fig:ext-5-closeup}).

We lock the laser to the cavity using the Pound-Drever-Hall scheme, shown in Fig.~\ref{fig:ext-1-setup}a. EOM~1 (iXblue, PHT MPZ-LN-10-00-P-P-FA-FA) generates the locking sidebands and is used together with acousto-optic modulator AOM~1 (G\&H, T-M200-0.1C2J-3F2P) to shift the locking frequency by one free spectral range of the cavity. This minimizes interference between the locking and the cooling light in detection.

To detect the particle motion in all directions, we use a heterodyne scheme, which mixes the scattered light with a local oscillator (LO). This enhances the signal interferometrically and shifts the signal to a spectral range of lower noise. The LO is blue-shifted by \qty{4.99814}{MHz} with respect to the tweezer beam using two polarization-maintaining fiber modulators (G\&H, T-M200-0.1C2J-3F2P): AOM~2 at \qty{197.5}{MHz} and AOM~3 at \qty{-202.49}{MHz} (Fig.~\ref{fig:ext-1-setup}c).

The scattered light transmitted by the cavity mirror is divided into its horizontal and vertical polarization components. They are individually combined with the LO beam using a $50:50$ fiber beam splitter (Thorlabs PN1550R5A2). Each polarization output is then detected by a balanced photodiode (Thorlabs PDB425C-AC). In both back-plane detections, we use variable-ratio fiber beam splitters (KS Photonics) to balance the outputs, which are also detected by balanced photodiodes (Thorlabs PDB440C-AC) (Fig.~\ref{fig:ext-1-setup}c).

After the optical trap, we collimate the tweezer light using a low-NA aspheric lens (Thorlabs C560TME-C) and isolate the particle signal using a split detection scheme (Fig.~\ref{fig:ext-1-setup}d). We use a D-shaped mirror to split the optical beam into two halves, which are detected by balanced photodiodes (Thorlabs PDB440C-AC). This detection is built for both the $x$- and the $y$-axis.

\subsection{Phase noise reduction}
In the presence of the cavity, laser phase noise can heat the mechanical motion~\cite{Meyer2019}. The cavity delays the release of the scattered light, effectively creating an unbalanced interferometer in heterodyne detection between scattered light and the local oscillator. The laser phase noise appears in cavity transmission as an increased noise background around the cavity mode resonance. In Fig.~\ref{fig:1-setup_and_two-mode_cooling}b, this is shown at a frequency around $\sim800\,\text{kHz}$ and fitted with a Lorentzian to extract the exact frequency and to determine the birefringence splitting. We also use the fitted frequency to determine the actual cavity-tweezer detuning and its error during the detuning scan in Fig.~\ref{fig:2}e. 

\begin{figure}[htbp]
    \centering
        \includegraphics[width=1\linewidth]{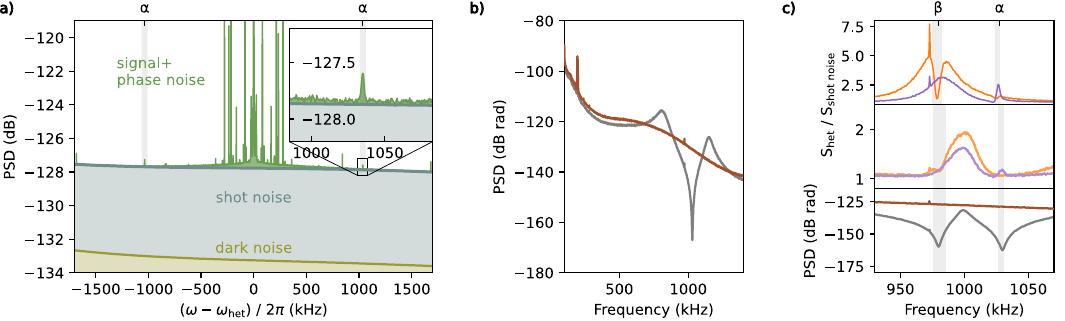}
    \caption{{\bf Characterization of the principal noise components.} {\bf a)} Power spectral density of the particle motion, as measured by back-scattering heterodyne detection \circled{1}. The $\alpha$-libration appears at a frequency of $2\pi\times 1030\,\text{kHz}$ and albeit rather weak, it can be clearly discerned because most noise components appear at lower frequency. The strong peaks below 400\,kHz are due to the linear motion along the three linear degrees of freedom, $x,y$ and $z$, as well as some higher harmonics and beat notes. Blocking all laser light provides the detector and oscilloscope dark noise. Blocking the signal and thus detecting the local oscillator provides the shot noise level. {\bf b)} Measured phase noise curve with an unbalanced Mach-Zehnder interferometer. The brown line shows the measured phase noise for relevant particle frequencies. With active feedback on EOM~2, the phase noise is reduced around the $\alpha$-libration by 35\,dB. This is crucial for achieving cooling the nanorotor deep into its 1D librational quantum ground state. {\bf c)} Dual-mode phase noise reduction for 2D ground-state cooling. In cavity transmission (output of detectors \circled{2} and \circled{3}), we observe the distortion of the motional peaksto a dip in the signal by noise squashing (top panel). When reducing the phase noise around the motional frequencies the phase noise pile-up is suppressed and the motional peaks become visible (center panel). We achieve a phase noise reduction of $30\,\text{dB}$ around the $\alpha$- and $\beta$-mode with a frequency separation of $2\pi\times 50\,\text{kHz}$ simultaneously (bottom panel), which enables ground-state cooling of both modes separately, as well as cooling close to the 2D ground state.
    }
    \label{fig:ext-2-noise}
\end{figure}

Strong cooling of the librational modes without active suppression of phase noise leads to noise squashing~\cite{safavi-naeini_laser_2013} (top panel of Fig.~\ref{fig:ext-2-noise}c), which distorts the motional sideband and generates a dip in the phase noise background. This also prevents accurate sideband thermometry.
We therefore implement a phase noise compensation scheme, using an unbalanced Mach-Zehnder interferometer \cite{parniak_high-frequency_2021} (Fig.~\ref{fig:ext-1-setup}b). The short arm contains a polarization-maintaining fiber attenuator to equalize the optical power in both arms. The long arm consists of a \qty{100}{m} single-mode fiber (SMF-28), enclosed in a chamber at prevacuum. This arm also includes a fiber stretcher to stabilize slow path length fluctuations ($>10\,\text{ms}$), and it combines a manual fiber polarization controller and a fiber PBS to correct for polarization changes. Light from both arms is recombined using a 50:50 fiber coupler and directed to a balanced detector (Thorlabs PDB450C-AC). After filtering, the interferometer output is fed back into EOM~2, which controls the phase of the tweezer light.

With active feedback, the noise level is reduced by more than $30\,\text{dB}$ both at a single frequency, seen in Fig.~\ref{fig:ext-2-noise}b, and two frequencies, shown in Fig.~\ref{fig:ext-2-noise}c, bottom. The reduction is also visible in cavity transmission, restoring the expected shape of the motional sidebands (center panel of Fig.~\ref{fig:ext-2-noise}).

\subsection{Theoretical description}
The nanorotor is an asymmetric rigid body ($I_c<I_b<I_a$), whose orientation in the lab frame $({\bf e}_x, {\bf e}_y, {\bf e}_z)$ is specified by the three Euler angles $(\alpha,\beta, \gamma)$, using $z$-$y'$-$z''$ convention (Inset Fig.~\ref{fig:1-setup_and_two-mode_cooling}a). 
Its optical response is characterized by the susceptibilities $\chi_a < \chi_b < \chi_c$ ~\cite{rudolph_theory_2021} which can be combined into the susceptibility tensor $\chi=\chi_a{\bf n}_1 \otimes {\bf n}_1+\chi_b{\bf n}_2 \otimes {\bf n}_2+\chi_c{\bf n}_3 \otimes {\bf n}_3$, where ${\bf n}_1,{\bf n}_2, {\bf n}_3$ are the basis vectors. 
The particle is illuminated by the linearly polarized tweezer field ${\bf E}_{\rm tw}({\bf r}) = E_{\rm tw}({\bf r}) {\bf e}_\phi$ of wavelength $2\pi/k$ with the tweezer mode amplitude $E_{\rm tw}({\bf r})\propto e^{i k z}$ propagating in ${\bf e}_z$ direction and the polarization direction ${\bf e}_\phi = {\bf e}_x \cos \phi + {\bf e}_y \sin \phi$. Coherent scattering of tweezer photons couples the deeply trapped particle rotations to two orthogonally polarized modes of the cavity field ${\bf E}_{\rm c}({\bf r}) = E_{\rm c}({\bf r}) ({\bf e}_y a_y + {\bf e}_z a_z)$, with $E_{\rm c}({\bf r}) \propto \cos(kx)$ denoting the cavity mode amplitude and $a_{y,z}$ the corresponding complex mode variables. The resulting interaction potential can be derived from the Lorentz torque acting on the particle as \cite{stickler_rotranslational_2016,rudolph_theory_2021,schafer_cooling_2021}
\begin{align}\label{eq:potentialfull}
    U = & -\frac{\varepsilon_0 V}{4} {\bf E}_{\rm tw} \cdot \chi {\bf E}_{\rm tw}^* \nonumber \\
    & - \frac{\varepsilon_0 V}{4} \left ( {\bf E}_c \cdot \chi {\bf E}_{\rm tw}^* + {\rm h.c.}\right ).
\end{align}
Here, $V$ denotes the particle volume and ${\bf R}$ is the particle center-of-mass position. Since the particle remains stably trapped at ${\bf R} \simeq 0$, the first term describes the librational trapping near $(\alpha,\beta) \simeq (\phi,\pi/2)$. The second term describes coupling of librations in $\alpha$ and $\beta$ to two orthogonally polarized cavity modes as well as trapping of $\gamma$. In our experiment, $\gamma \simeq 0$ or $\gamma \simeq \pi/2$ because the cavity modes are polarized along ${\bf e}_y$ and ${\bf e}_z$. 
In the following, we assume $\gamma \simeq \pi/2$, the case $\gamma \simeq 0$ can be obtained through the replacement of indices $a\leftrightarrow b$. The librational frequencies for deviations of $\alpha$ and $\beta$ from their equilibrium orientation are given by:
\begin{align}\label{eq:librationalfrequencies}
\begin{split}
    \Omega_\alpha=\sqrt{\frac{\varepsilon_0V}{2I_b}(\chi_c-\chi_a)}|E_{\rm tw}(0)|\\ \Omega_\beta=\sqrt{\frac{\varepsilon_0V}{2I_a}(\chi_c-\chi_b)}|E_{\rm tw}(0)|
\end{split}
\end{align}

The second term in Eq.~\eqref{eq:potentialfull} decomposes into an orientation-independent term that drives the in-plane cavity mode $a_y$ and an orientation-dependent term that describes coupling between the cavity modes and the particle librations. Specifically, the former term can be written in the form $V_{\rm dr} = \hbar (\eta a_y + {\rm h.c.})$ with the pump rate:
\begin{equation}
    \eta = -\frac{\varepsilon_0 \chi_a V}{4 \hbar} E_{\rm c}(0) E_{\rm tw}^*(0) \sin \phi
\end{equation}

Likewise, the coupling between librations and the cavity modes follows from the orientation of the susceptibility tensor.
For $\phi = 0$, the coupling becomes approximately linear in both librational degrees of freedom:
\begin{equation}\label{eq:alphabetacoupling}
    U_\text{int}\approx k_\alpha\alpha a_y+k_\beta\beta a_z+{\rm h.c.},
\end{equation}
where the complex-valued constants for both librational modes are given by:

\begin{align}\label{eq:librationalkonstant}
\begin{split}
    k_\alpha = \frac{\varepsilon_0 V}{4} (\chi_c-\chi_a) E_c(0) E^*_{\rm tw}(0)\\
    k_\beta= \frac{\varepsilon_0 V}{4} (\chi_c-\chi_b) E_c(0) E^*_{\rm tw}(0)
\end{split}
\end{align}

For both values of $\gamma$, Eq.~(\ref{eq:alphabetacoupling}) shows that $\alpha$ couples to the in-plane cavity mode $a_y$, while $\beta$ couples to the out-of-plane cavity mode $a_z$. Cavity-transmission spectra (Fig.~\ref{fig:1-setup_and_two-mode_cooling}b) consistently show $\Omega_\alpha > \Omega_\beta$ across all nanorotors trapped in our setup, which is compatible with $\gamma \simeq \pi/2$ and motivates this choice in our modeling.

We define the librational mode variables $b_\alpha = \alpha_{\rm zpf} (\alpha + i p_\alpha/I_\alpha \Omega_\alpha)$ and $b_\beta = \beta_{\rm zpf} (\beta - \pi/2 + p_\beta/I_\beta \Omega_\beta)$, with zero-point amplitudes ${\alpha}_{\rm zpf} = \sqrt{\hbar/2I_{b}\Omega_{\alpha}}$ and ${\beta}_{\rm zpf} = \sqrt{\hbar/2I_{a}\Omega_{\beta}}$, to obtain the quantized interaction Hamiltonian~(\ref{eq:quantinteraction_hamiltonian}), where we introduced the coupling rates $g_\alpha = \alpha_\text{zpf} k_\alpha$ and $g_\beta = \beta_\text{zpf} k_\beta$.
In summary, this leads to the total libration-cavity Hamiltonian~(\ref{eq:hamiltonian}).
A standard calculation then yields the optomechanical damping rates and the resulting steady-state occupation \eqref{eq:final_occupation} \cite{rudolph_theory_2021}.

\subsection{Optomechanical coupling}
The optomechanical coupling determines the interaction between the particle and cavity mode and thereby the cooling performance. By solving the equations of motion, with cooling providing additional damping, we obtain an effective motional linewidth
\begin{align}
\gamma_\mu^{\text{eff}}(\omega) &= 
\gamma_\mu + \frac{4 |g_\mu|^2 \Omega_\mu \Delta_c \kappa}
{\left[\left( \tfrac{\kappa}{2} \right)^2 + (\omega + \Delta_c)^2 \right] \left[ \left( \tfrac{\kappa}{2} \right)^2 + (\omega - \Delta_c)^2 \right]},
\end{align}
which depends on the coupling strength.
In the regime of strong cooling, when the cavity resonance is close to the mechanical frequency, energy loss through the cavity determines the damping and the cavity-induced linewidth dominates over the thermal linewidth $\gamma_\mu$. We use this expression to fit the linewidths extracted from cavity-detuning scans for 1D (Fig.~\ref{fig:2}d) and 2D (Fig.~\ref{fig:ext-6-2D}c) cooling with a constant coupling. We verify the extracted coupling by additionally fitting the observed optical spring effect (Fig.~\ref{fig:2}c):
\begin{align}
\Omega_\mu^{\mathrm{eff}}(\omega)
= \sqrt{
\Omega_\mu^2
- \frac{4\,|g_\mu|^2\,\Omega_\mu\,\Delta_c
\left[\left(\tfrac{\kappa}{2}\right)^2 - \omega^2 + \Delta_c^2\right]}
{\left[\left(\tfrac{\kappa}{2}\right)^2 + (\omega + \Delta_c)^2\right]
 \left[\left(\tfrac{\kappa}{2}\right)^2 + (\omega - \Delta_c)^2\right]}
}
\end{align}
Since the optomechanical coupling is determined by the rotor geometry, we can determine the moment of inertia for each mode. Combining Equations~(\ref{eq:librationalfrequencies}) and (\ref{eq:librationalkonstant}) with the zero-point fluctuation, we calculate:
\begin{align}
    I_b=\frac{|g_\alpha|^2}{\Omega_\alpha^3}\frac{|E_\text{tw}(0)|^2}
    {|E_c(0)|^2 8\hbar},\quad
    I_a=\frac{|g_\beta|^2}{\Omega_\beta^3}\frac{|E_\text{tw}(0)|^2}{|E_c(0)|^2 8\hbar}
\end{align}

\subsection{Noise contributions}
\label{meth:calc_of_occ}
For quantum-limited measurements, the signal must be isolated from noise. The noise contributions in backscattering detection are shown in Fig.~\ref{fig:ext-2-noise}a. The raw spectrum contains dark noise (photodetector and oscilloscope), shot noise, and phase noise of the local oscillator. The latter originates from the frequency generators that drive the local oscillator AOMs~2 and 3 (Fig.~\ref{fig:ext-1-setup}c). In postprocessing, we therefore subtract the background levels, extracted from the Lorentzian fits. Additionally, the detector sensitivity shows a weak frequency dependence, which differs for the Stokes and anti-Stokes peaks. The sensitivity is calibrated by acquiring spectra of the dark noise and of the local oscillator's shot noise. Since shot noise is white, any residual frequency dependence must be due to the detector response. We therefore divide the background-corrected signals by the difference between shot noise and dark noise.

\subsection{Occupation number}
The areas of the Stokes and anti-Stokes peak scale with the occupation number $n$ as $A_{\rm S}=C~(n+1)$ and, respectively, $A_{\rm aS}=C~n$. Thus, the occupation can be extracted by
taking the ratio of the Stokes and anti-Stokes peak areas. This works well for occupation numbers close to the ground state, but yields large uncertainties for $n \gg 1$, where the ratio is close to $1$. Here, we use the fact that the difference of the sideband areas $A_{\mathrm{S}} - A_{\mathrm{aS}} = C$ is independent of the occupation $n$. Therefore, we determine $C$ by averaging the differences of areas for all spectra in a series of measurements and subsequently extract the occupation number for a specific detuning $n = (A_{\mathrm{S}} + A_{\mathrm{aS}} - C) / 2C$. We normalize the peak area with $C$ for both peaks, such that the difference of both yields $1$. The resulting normalized sideband areas are shown in Fig.~\ref{fig:2}b for 1D cooling and in Fig.~\ref{fig:ext-6-2D}a--b for 2D cooling.

\subsection{Heating rates}
In the absence of external heating, cooling is governed by the cavity-enhanced imbalance between anti-Stokes and Stokes scattering. For both processes, we define the weak-coupling damping and heating rate as
\begin{equation}
    A_\mu^\pm = \frac{|g_\mu|^2 \kappa}{(\kappa/2)^2 + (\Delta \pm \Omega_\mu)^2},
   \label{eq:optomechanical-damping-rates} 
\end{equation}
which yields, together with Equation (\ref{eq:final_occupation}), a minimum occupation number of $n_\text{min}=\kappa^2/4\Omega^2$. It depends only on the cavity linewidth and the mechanical frequency. 
For librational frequencies of $\sim2\pi\times 1\,\text{MHz}$ this implies a theoretical lower bound of $n_\alpha\approx n_\beta\approx 2.5\times 10^{-4}$, far below our measured values. The system must therefore be limited by other sources, such as recoil heating, gas collisions, or phase noise. 

The recoil limit depends on both the cavity and tweezer parameters. For our linearly polarized tweezer we estimate $\Gamma^\text{recoil}=3.2\,\text{kHz}$~\cite{rudolph_theory_2021}, which limits cooling to $n_\text{recoil}=0.064$. 
Phase noise and collisional contributions, however, vary with the particle geometry, as this determines the librational frequency and the collisional cross section. The phase noise occupation can be obtained from Eq.~(\ref{eq:final_occupation}).

We analyze heating and decoherence for the ground state cooled nano-cluster (Fig.~\ref{fig:2}) and find from the frequency dependence of the occupation
that the phase noise contribution of $n_\phi(\Omega_\alpha)=0.00^{+0.01}$ is negligible. Additionally, the fit reveals a total heating rate of $\Gamma_\alpha=6.8\pm0.7\,\text{kHz}$, originating from both recoil and thermal noise. Since the former is pressure-independent, the thermal part follows by subtraction from the total heating rate $\Gamma_\alpha^\text{thermal}=3.6\pm0.8\,\text{kHz}$. For this cluster particle, recoil and thermal heating contribute approximately equally.

The same noise analysis can be performed for the trapped nano-dumbbell, where we treat both librational modes separately (Fig.~\ref{fig:3}). For the $\beta$-libration, the fit finds the phase noise to dominate with an occupation of $n_\phi(\Omega_\beta)=0.38\pm0.17$, while the $\alpha$ mode is again only barely affected by it, with $n_\phi(\Omega_\alpha)=0.00^ {+0.07}$.
From the fitted total heating rates in both dimensions, $\Gamma_\beta=20\pm4\,\text{kHz}$ and $\Gamma_\alpha=18\pm2\,\text{kHz}$, we estimate the thermal heating rates $\Gamma_\beta^\text{thermal}=16\pm4\,\text{kHz}$ and $\Gamma_\alpha^\text{thermal}=14\pm2\,\text{kHz}$, respectively. 
We conclude that collisional heating dominates the $\alpha$-mode, while the $\beta$-libration is also limited by phase noise.

\end{document}

%% file: authors.tex
\author{Stephan Troyer}
\affiliation{University of Vienna, Faculty of Physics \& Vienna Doctoral School in Physics \& Vienna Center for Quantum, Boltzmanngasse 5, 1090 Vienna, Austria}

\author{Florian Fechtel}
\affiliation{University of Vienna, Faculty of Physics \& Vienna Doctoral School in Physics \& Vienna Center for Quantum, Boltzmanngasse 5, 1090 Vienna, Austria}

\author{Lorenz Hummer}
\affiliation{University of Vienna, Faculty of Physics \& Vienna Doctoral School in Physics \& Vienna Center for Quantum, Boltzmanngasse 5, 1090 Vienna, Austria}

\author{Henning Rudolph}
\affiliation{University of Duisburg-Essen, Faculty of Physics, Lotharstra\ss e 1, 47057 Duisburg, Germany}

\author{Benjamin A. Stickler}
\email{benjamin.stickler@uni-ulm.de}
\affiliation{Institute for Complex Quantum Systems and Center for Integrated Quantum Science and Technology, Ulm University, Albert-Einstein-Allee 11, 89069 Ulm, Germany}

\author{Uro\v {s} Deli\' c}
\affiliation{University of Vienna, Faculty of Physics \& Vienna Doctoral School in Physics \& Vienna Center for Quantum, Boltzmanngasse 5, 1090 Vienna, Austria}
\affiliation{Vienna Center for Quantum Science and Technology, Atominstitut, TU Wien, Stadionallee 2, 1020 Vienna, Austria}

\author{Markus Arndt}
\email{markus.arndt@univie.ac.at}
\affiliation{University of Vienna, Faculty of Physics \& Vienna Doctoral School in Physics \& Vienna Center for Quantum, Boltzmanngasse 5, 1090 Vienna, Austria}